\documentclass[aps,prd,reprint,amsmath,amssymb,showpacs,superscriptaddress]{revtex4-1}
\usepackage{graphicx}
\usepackage{subfigure}
\usepackage{dcolumn}
\usepackage{bm}
\usepackage{slashed}
\usepackage{ulem}
\usepackage{epstopdf}
\usepackage{mathrsfs}
\usepackage{color,xcolor}

\usepackage{hyperref}
\newcommand{\td}{\ensuremath{\textrm{d}}}
\newcommand{\trD}{\ensuremath{\textrm{tr}_{\textrm{D}}}}
\newcommand{\Tr}{\ensuremath{\textrm{Tr}}}

\begin{document}
\title{New insight on the quark condensate beyond chiral limit}

\author{Ling-feng Chen}
\email{lfchen1130@pku.edu.cn}
\affiliation{Department of Physics and State Key Laboratory of Nuclear Physics and Technology, Peking University, Beijing 100871, China.}

\author{Zhan Bai}
\email{baizhan@pku.edu.cn}
\affiliation{Department of Physics and State Key Laboratory of Nuclear Physics and Technology, Peking University, Beijing 100871, China.}

\author{Fei Gao}
\email{f.gao@thphys.uni-heidelberg.de}
\affiliation{Institut f\"{u}r Theoretische Physik, Universit\"{a}t Heidelberg, Philosophenweg 16, 69120 Heidelberg, Germany.}

\author{Yu-xin Liu}
\email{yxliu@pku.edu.cn}
\affiliation{Department of Physics and State Key Laboratory of Nuclear Physics and Technology, Peking University, Beijing 100871, China.}
\affiliation{Collaborative Innovation Center of Quantum Matter, Beijing 100871, China.}
\affiliation{Center for High Energy Physics, Peking University, Beijing 100871, China.}

\date{\today}

\begin{abstract}
With analyzing the mass function obtained by solving Dyson-Schwinger Equations,
we propose  a cut-off independent  definition of quark condensate beyond chiral limit.
With this well-defined condensate, we then  analyze the evolution of the condensate and its  susceptibility with the current quark mass.
The susceptibility shows a critical mass  in the neighborhood of the s-quark current mass, which defines a transition boundary for internal hadron dynamics.
\end{abstract}

\maketitle

\section{Introduction}
\label{intro}

Chiral symmetry and its breaking play significant roles in QCD phase structure as well as hadron dynamics.
The current quark mass is about $3\sim 5\;$MeV,
but a constituent quark inside a proton acquires a mass of about $300\sim 500\;$MeV through interaction.
This effect is called dynamical chiral symmetry breaking (DCSB) and is essential in the study of QCD properties
(for a recent review see Ref.~\cite{Roberts:2021arXiv}.
See also Refs.~\cite{Hawes:1994PRD,Munczek:1995PRD,Chang:2007PRC,Williams:2007PLB,Chen:2009JPG,Aguilar:2011PRD,Mitter:2015PRD,Braun:2016PRD,Abhishek:2019PRD,Roberts:2020Symmetry}).
In order to describe the transition from a dynamical chiral symmetric (DCS) phase to a DCSB phase,
the chiral condensate, {\it i.e.}, the expectation value of the composite operator $\bar{q}q$,
is usually applied as the order parameter~\cite{Genon:2004,Fischer:2011PLB,Brodsky:2012PRC,HDTrottier:2013,Cui:2015AP,Raval:2019NPB,Braun:2020PRD,Gao:2020PRD,Xu:2020PRD,Bai:2020arXiv,Gunkel:2020arXiv},
and the chiral condensate is also related to many important problems such as the pion-nucleon sigmaterm~\cite{Huang:2020PRD},
	the cosmological constant~\cite{Brodsky:2011PNAS},
	and thermodynamic quantities~\cite{Isserstedt:2020arXiv}.

The chiral condensate and DCSB has been extensively studied in lattice QCD~\cite{lattice1,lattice2,ABazavov,Bazavov:2012PRD},
functional renormalization group methods~\cite{FRG1,FRG2,FRG3,FRG4}, Dyson-Schwinger equations(DSEs)~\cite{Roberts0,Roberts1,Roberts2,CLR,Papavassiliou,FeiGaos,sxqins,Fischers} and  effective field models
such as the (Polyakov improved) Nambu--Jona-Lasinio model~\cite{NJL,Klevansky:1992,Buballa:2005PR,Ratti:2006PRD,Abhishek:2019PRD},
	 and quark meson model~\cite{Schaefer:2007PRD,Zacchi:2015PRD,FRG3,Li:2019PRD}.
Among these theoretical approaches, DSEs approach is a first-principle, non-perturbative continuum method.
It is able to deal with the DCSB and confinement simultaneously,
and has been widely applied to study the hadron properties and the QCD phase transitions.

However, despite the great success of the theoretical studies, there is still ambiguity in the definition of the condensate,
especially when quark has a non-zero current mass, {\it i.e.}, the explicitly chiral symmetry breaking (ECSB) is considered.
The ECSB comes from the Higgs mechanism, which is significantly different from the DCSB mechanism in QCD.
The effect of ECSB and its interference on DCSB is important for understanding the different mass generation mechanisms and their relations on the properties of quarks and hadrons
~\cite{Roberts:2020Symmetry,interference2,interference3,interference4}.
However, on one hand, it is  challenging  to well define the chiral condensate due to the quadratic divergence brought by the current quark mass term
~\cite{MRoberts:1997,HDTrottier:2013,Roberts:2010,Roberts:2003}.
On the other hand, it is in principle difficult to separate the DCSB effect from the explicit mass scale from ECSB term.

There are several different ways to remove the quadratic divergence of the condensate beyond chiral limit.
For example, it has been proposed that the divergence can be canceled by taking trace of the sum of different DSE solutions~\cite{Chang:2007PRC,Williams:2007PLB,Chen:2009JPG}.
However, at high temperature or chemical potential or current quark mass, there might be only one solution and this method cannot be adapted there.

The most straight forward way is to eliminate the quadratic divergence by doing subtraction.
One of the commonly used scheme is to subtract a fraction of strange quark condensate from light quark condensate,
    where the fraction coefficient is carefully designed to remove the quadratic divergence~\cite{Fischer:2019PPNP,Braun:2020PRD,Bazavov:2012PRD}.
Another scheme is to subtract the derivative of the condensate~\cite{FeiGao:2016}.
Both of these two schemes is successful in removing the quadratic divergence induced by the non-zero current quark mass,
     but there is still a logarithmic divergence, which already exists at  chiral limit.

In Ref.~\cite{Williams:2007PLB}, it has also been proposed that the condensate can be extracted from the ultraviolet behavior of the mass function.
However, in case of finite current quark mass, the ultraviolet contribution to the condensate is extremely smaller than that from the current quark mass,
and the fitting process is the not reliable enough.
%

Inspired by these earlier works,
we find that if we take the first subtraction scheme mentioned above to eliminate the contribution of the current quark mass,
   and then fit the ultraviolet behavior of the condensate,
 we can extract the quark condensate from quark propagator while avoiding both the logarithm and the quadratic divergences,
   and this process can be extended to large current quark mass since it does not require the existence of multi-solutions of the DSEs,
{\it i.e.}, we verify a well-defined and divergence-free quark condensate beyond the chiral limit directly in terms of  dressed quark propagator.

In our calculation, the current quark mass dependence of the condensate reveals a critical mass in the neighborhood of the s-quark current mass.
The critical mass confirms the previous studies of parton distribution amplitude (PDA) of mesons,
where a critical mass has also been found near s-quark mass that the respective PDA becomes asymptotic~\cite{PDA1,PDA2}.
These studies indicates a transition boundary for internal hadron dynamics between the different mass generation mechanisms.


The remainder paper is organized as follows.
In Section~\ref{Gap}, we describe the DSE approach and the setups.
In Section~\ref{Cond}, we describe the theoretical framework of extracting the condensate introduced herein.
In Section~\ref{RaD}, we present our numerical results and discussions.
Section~\ref{SaO} provides a summary and perspective.

\section{The Gap Equation} \label{Gap}
\begin{figure}[!htp]
\centering
\includegraphics[width=0.44\textwidth]{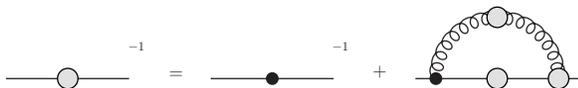}
\caption{The Feynmann diagram for the Dyson-Schwinger equation of the quark propagator in
Eq.~\eqref{eq:DSE}. The solid line with gray circle denotes the dressed quark propagator, the solid line with black circle denotes the bare quark propagator,
and the curly line with gray circle denotes the dressed gluon propagator, and the black and gray circles stands for the bare vertex, the dressed vertex, respectively.}
\label{fig:gap}
\end{figure}

The quark condensation is the trace of the fully dressed quark propagator.
The starting point is the renormalized Dyson-Schwinger equation for the quark propagator $S$, as schematically depicted in Fig.~\ref{fig:gap}.
It reads
\begin{equation}\label{eq:DSE}
S^{-1}(p)=Z_2(i\slashed{p}+Z_m m^{\zeta})+\Sigma(p),
\end{equation}
with the self energy
\begin{equation}\label{eq:DSE2}
\Sigma(p)= g^2 Z_1\int_q^{\Lambda} D_{\mu\nu}(p-q)  \frac{\lambda^a}{2}\gamma_{\mu}S(q) \frac{\lambda^a}{2}\Gamma_{\nu}(q,p),
\end{equation}
where $Z_1$, $Z_2$, and $Z_m$ are the vertex, quark wave-function, and mass renormalization constants, respectively.
$m^{\zeta}$ is the renormalized current quark mass,
${\lambda^a}$ are the color matrices,
$\int_q^{\Lambda}$ represents a Poincar\'e invariant regularization of the four-dimensional integral,
with $\Lambda$ the ultraviolet regularization mass scale.
$\Gamma_{\nu}$ and $D_{\mu\nu}$ are the dressed quark-gluon vertex and the dressed gluon propagator, respectively.

From the above equations, we can see that the propagator $S$ depends on the dressed gluon propagator $D_{\mu\nu}$ and the dressed quark-gluon vertex $\Gamma_{\nu}$ explicitly,
which must be specified.
Instead of solving the coupled quark, ghost and gluon system, one may rather choose to employ some suitable {\it Ans\"atz} for the coupling and interaction in Eq.\ \eqref{eq:DSE2},
which has sufficient integrated strength in the infrared to achieve the dynamical mass generation mechanism.

The commonly used {\it Ans\"atz} for the $g^2D_{\mu\nu}$ is:
\begin{align}
g^2 D_{\mu\nu}(k) = k^2 \mathcal{G}(k^2)D_{\mu\nu}^{\text{free}}(k),\label{InteractionModel}
\end{align}
where $D_{\mu\nu}^{\text{free}}(k)=(\delta_{\mu\nu}-\frac{k_{\mu}k_{\nu}}{k^2})\frac{1}{k^2}$ is the Landau-gauge free gluon propagator.
The interaction model $\mathcal{G}(k^2)$ is written as
\begin{align}
k^2 \mathcal{G}(k^2) = k^2\mathcal{G}_{\mathrm{IR}}(k^2)+4\pi\tilde{\alpha}_{\mathrm{pQCD}}(k^2)\,,
\end{align}
where $\tilde{\alpha}_{\mathrm{pQCD}}(k^2)$ is a bounded and monotonically decreasing regular continuation of the perturbative QCD running coupling to all values of spacelike-$k^2$,
and $\mathcal{G}_{\mathrm{IR}}(k^2)$ is the interaction at infrared region and dominates in the region $|k|<\Lambda_{\mathrm{QCD}}$.
The form of $\mathcal{G}_{\mathrm{IR}}(k^2)$ determines whether the DCSB and/or confinement can be realized.

In this work we adopt the Qin-Chang (QC) model~\cite{QC}, and the interaction is expressed as ($s=k^2$):
\begin{equation}\label{QC}
\mathcal{G}(s)=\frac{8\pi^2}{\omega^4}D e^{-s/\omega^2}+\frac{8\pi^2\gamma_m\mathcal{F}(s)}{\ln[\tau+(1+s/\Lambda_{\mathrm{QCD}}^2)^2]},
\end{equation}
where $\mathcal{F}(s)=(1-e^{-s/4m_{t}^{2}})/s$ with $m_{t}=0.5\,$GeV,
	  $\gamma_m=12/(33 - 2 N_{f})$ is the dimension anomaly with the flavor number $N_{f}$,
	 and $\tau=e^{2} -1$ is a constant.
Following Ref.~\cite{sxqin:2018}, we take $N_{f} =4$ and $\Lambda_{\mathrm{QCD}}=0.234\,$GeV.
The parameters $D$ and $\omega$ control the strength and the width of the interaction, respectively.
In fact, observable properties of light-quark ground-state vector- and isospin-nonzero pseudoscalar mesons are insensitive to variations of $\omega \in[0.4,0.6]\,$GeV,
as long as
%
$$ \varsigma^3:=D\omega = {\rm constant}\,. $$
%
In this work, following the commonly used values, we set $D=1.024$ GeV$^2$ and $\omega=0.5$ GeV \cite{sxqin:2018x}.
Since this model assumes a rainbow vertex truncation,
%
$ \Gamma_{\mu}(q,p)=\gamma_{\mu} \, , $
%
the solutions are not multiplicatively renormalizable and so depend on the chosen renormalization point.
The renormalization scheme is one of modified momentum subtraction at some point $\zeta$. We set  $\zeta=\Lambda$,
which means that we do the renormalization at infinity, and all the renormalization constants becomes $1$.

The solution of the gap equation, {\it i.e.}, the dressed quark propagator, can be decomposed as
\begin{eqnarray}
S^{-1}(p)&=&i \slashed{p} A(p^2) +B(p^2),    \label{quark_lorentz2}
\end{eqnarray}
with the momentum subtraction renormalization condition
\begin{equation}
S^{-1}(p)|_{p^2=\zeta^2}=i\slashed{\zeta} + m^\zeta,
\end{equation}
where $\zeta$ is again the renormalization point and $m^\zeta$ the renormalized current quark mass. The dressed quark mass function is defined as
\begin{equation}\label{MassFunc}
 M(p^2)=B(p^2,\zeta^2)/A(p^2,\zeta^2),
\end{equation}
which is independent of the renormalization point $\zeta$. Note that $m^\zeta$ is exactly the mass function evaluated at the renormalization point, {\it i.e.}, $m^\zeta=M(\zeta^2)$.

\section{RESULTS AND DISCUSSIONS}

\subsection{Quark condensate beyond chiral limit}
\label{Cond}

Now applying the conventional definition of chiral condensate as the trace of quark propagator, one can compute the condensate via:
\begin{equation} \label{eq:trace}
\begin{split}
\Tr [S]=&-N_cN_f \int_{k}^{\Lambda}\trD[S(k, \zeta)]\\
=&-\frac{N_{c}N_{f}}{2\pi^2}\int^{\Lambda}_0\frac{k^3\td k M(k^2,\zeta^2)/A(k^2,\zeta^2)}{k^2+M^2(k^2,\zeta^2)}.
\end{split}
\end{equation}

According to the operator product expansion (OPE) technique, the asymptotic behavior of the mass function at large momentum is~\cite{Roberts0,Williams:2007PLB}:
\begin{equation}\label{eq:masstail}
\begin{split}
M(p^2)_{\rm asym} =
\frac{\mathcal{C}}{p^2}&\left[\frac{1}{2}\ln(p^2/\Lambda_{\rm QCD}^2)\right]^{\gamma_m -1}\\
&+\hat{m}\left[\ln(p^2/\Lambda_{\rm QCD}^2)\right]^{-\gamma_m}\,,
\end{split}
\end{equation}
where $\mathcal{C}=\frac{-2 \pi^2 \gamma_m}{3}\frac{\langle \bar{q}q\rangle^{\zeta 0}}{[\frac{1}{2}\ln(\zeta^2/\Lambda_{\rm QCD}^2)]^{\gamma_m}}$ and $\hat{m}=m^{\zeta}[\frac{1}{2}\ln(\zeta^2/\Lambda_{\rm QCD}^2)]^{\gamma_m}$ are both renormalization-independent quantities.
$\langle\bar{q}q\rangle^{\zeta 0}$ is the condensate at chiral limit with renormalization point $\zeta$.
By inserting Eq.~(\ref{eq:masstail}) into Eq.~(\ref{eq:trace}), we have:
\begin{equation}\label{eq:trs}
\begin{split}
\Tr[ S_{\rm asym}] =&-\langle \bar{q}q\rangle^{\zeta 0}\frac{[\ln(\Lambda_{}^2/\Lambda_{\rm QCD}^2)]^{\gamma_m}}{[\ln(\zeta^2/\Lambda_{\rm QCD}^2)]^{\gamma_m}}\\
&+\frac{N_{c}}{2\pi^2}\int^{\Lambda} p \hat{m} \left[ \ln{\big{(} p^{2}/{\Lambda^{2}_{\mathrm{QCD}}} \big{)}} \right]^{-\gamma_{m}} \td p \,.
\end{split}
\end{equation}
It is apparent that there exist a logarithmic divergence when $m^\zeta=0$,
and an extra quadratic divergence for finite $m^\zeta$.

From Eq.~(\ref{eq:trs}), one can notice that the trace of the quark propagator is cut-off dependent,
and needs a subtraction to eliminate the quadratic divergence that comes from the linear dependence of the current quark mass in the mass function,
and then a renormalization to deal with the logarithmic divergence.


In Refs.~\cite{HDTrottier:2013,FeiGao:2016}, the condensate beyond chiral limit is defined as:
\begin{equation} \label{eq:subtra}
-\langle \bar{q}q\rangle_{m^{\zeta}}:=\left(1-m^{\zeta}\frac{\partial}{\partial m_{\zeta}}\right)\Tr[ S].
\end{equation}
Applying Eq.~(\ref{eq:subtra}) on Eq.~(\ref{eq:trs}),
we find that the second term in Eq.~(\ref{eq:trs}) vanishes.
However the logarithmic divergence  remains as:
\begin{equation}\label{trs_sub}
\begin{split}
&\left(1-m^{\zeta}\frac{\partial}{\partial m^\zeta}\right)\Tr [S_{\rm asym}]\\
&=-\frac{\left(1-m^\zeta\frac{\partial}{\partial m^\zeta}\right)\langle \bar{q}q\rangle^{\zeta 0}_{}}{\left[\frac{1}{2}\ln(\zeta^2/\Lambda_{QCD}^2)\right]^{\gamma_m}}\left[\frac{1}{2}\ln(\Lambda_{}^2/\Lambda_{QCD}^2)\right]^{\gamma_m}.
\end{split}
\end{equation}
Therefore, we can define the renormalization and cut-off independent condensate $\langle\bar{q}q\rangle$ as:
\begin{equation}
\langle\bar{q}q\rangle=\left(1-m^{\zeta}\frac{\partial}{\partial m^{\zeta}}\right)\frac{\langle \bar{q}q\rangle^{\zeta 0}_{}}{\left[\frac{1}{2}\ln(\zeta^2/\Lambda_{QCD}^2)\right]^{\gamma_m}},
\end{equation}
and the r.h.s. of Eq.~(\ref{trs_sub}), which is the asymptotic behavior of the condensate, can be written as:
\begin{equation}\label{eq:fit_func}
\begin{split}
\mathcal{F}(\langle\bar{q}q\rangle,\Lambda_{\textrm{QCD}},\gamma_m)
=-\langle\bar{q}q\rangle\left[\frac{1}{2}\ln(\Lambda_{}^2/\Lambda_{QCD}^2)\right]^{\gamma_m}.
\end{split}
\end{equation}

After figuring out the asymptotic behavior of the subtracted condensate,
we can use $\mathcal{F}(\langle\bar{q}q\rangle,\Lambda_{\textrm{QCD}},\gamma_m)$ defined in Eq.(\ref{eq:fit_func}) to fit the trace of the propagator, $\left(1-m^{\zeta}\frac{\partial}{\partial m_{\zeta}}\right)\Tr[ S]$,
and we can extract the condensate $\langle\bar{q}q\rangle$ as the fitting parameter.
For simplicity, we do the renormalization at infinity, {\it i.e.}, taking $\zeta=\Lambda$,
and all the renormalization constants becomes $1$.
The renormalized condensate at renormalization point $\zeta$ reads:
\begin{equation}\label{eq:evolve}
\langle \bar{q}q\rangle_{\zeta}=\langle \bar{q}q\rangle\left[\frac{1}{2}\ln(\zeta^2/\Lambda_{\rm QCD}^2)\right]^{\gamma_{m}} \, .
\end{equation}

\begin{figure}[!htp]
\centering
\includegraphics[width=0.44\textwidth]{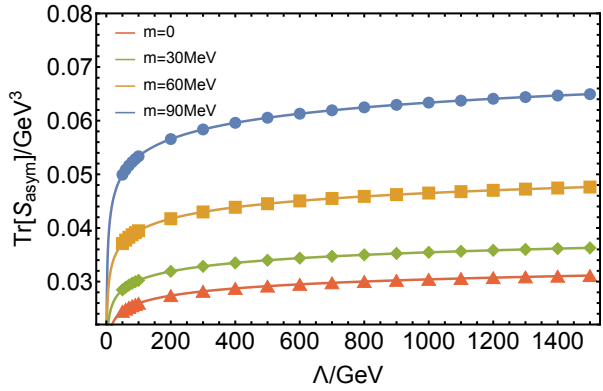}
\vspace*{-3mm}
\caption{(color online) The calculated trace of the quark propagator as a function of cut-off $\Lambda$.
The symbols with different shapes correspond to the trace of the propagator calculated from Eq.~(\ref{eq:trace}),
and the solid lines with different colors correspond to the fitted curves in Eq.~(\ref{eq:fit_func}).}
\label{fig:c0}
\end{figure}

The result of the fitting is  shown in Fig.~\ref{fig:c0}.
As can be seen from the figure, the fit is  with a quite high precision for a range of current quark masses.
In chiral limit, the fitting parameters are $\Lambda_{\rm QCD}=0.2749\;$GeV, $\gamma_{m}=0.4799$,
 and   $-\langle \bar{q}q\rangle^0_{}=  ( 223\;{\rm MeV} )^3$.
The condensate at 2GeV is derived from Eq.~(\ref{eq:evolve}): $-\langle \bar{q}q\rangle^0_{\zeta=2GeV}=( 249\;\textrm{MeV})^3$.
This is comparable with lattice studies (see, e.g., (252(5)(10)\; MeV)$^3$ in Ref.~\cite{lqcd0}, and other lattice results in Refs.~\cite{Fukaya1,Fukaya2,lqcd,lqcd1,lqcd2,lqcd3,lqcd4}).

\begin{figure}
\centering
\includegraphics[width=0.44\textwidth]{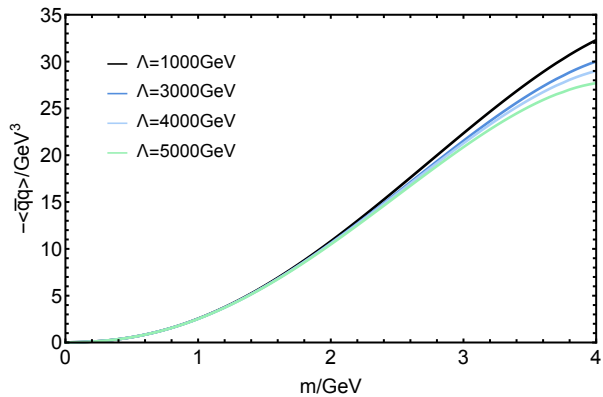}
\vspace*{-3mm}
\caption{(color online) The extracted current mass dependence of the condensate with different cut-offs from $\Lambda=1000$ GeV to  $\Lambda=5000$ GeV.}
\label{fig:cm}
\end{figure}

Now we get a cut-off independent chiral condensate in terms of the quark propagator beyond chiral limit.
As shown in Fig.~\ref{fig:cm}, with changing the cut-off from $\Lambda=1000\;$GeV to $\Lambda=5000\;$GeV,
the chiral condensate is cut-off independent within errors in a wide range of $m$ from 0 to 4 GeV.
For larger quark mass, the error increases slightly due to the sizable current quark mass scale that drives the ultraviolet behavior of quark mass function deviates from  the asymptotic form.
The error can be reduced by setting a larger cut-off.

\subsection{Interference between  DCSB and ECSB}
\label{RaD}

From Fig.~\ref{fig:cm}, we find that the chiral condensate increases monotonically with the current quark mass.
This behavior seemingly means that the DCSB effect is stronger for heavier quark.
This is in contradiction to the non-relativistic QCD (NRQCD) computation of heavy quarkonium,
whose result shows that the heavier system can be better described with current quark by the non-relativistic potential~\cite{NRQCD1,NRQCD2,NRQCD3}.
However, the increasing condensate only means that the mass scale brought by the chiral symmetry breaking effect grows along with the scale of current quark mass.
This mass scale contains three parts of contribution: the dynamical part (DCSB), the explicit part (ECSB) and their interference.
By doing the subtraction, the ECSB effect is excluded from the definition of condensate,
and the condensate is consist of the DCSB and the interference part.
The sum of the DCSB and the interference effect increases as the ECSB increases,
but the proportion from the DCSB effect should be small when the current quark mass is large,
since the interaction is negligible for extremely heavy quark system.
To illustrate this, we can define a dimensionless chiral susceptibility as:
\begin{equation}
\chi(m)=\frac{-\partial{\langle\bar{q}q\rangle^{1/3}}}{\partial m}.
\end{equation}

\begin{figure}
\centering
\includegraphics[width=0.44\textwidth]{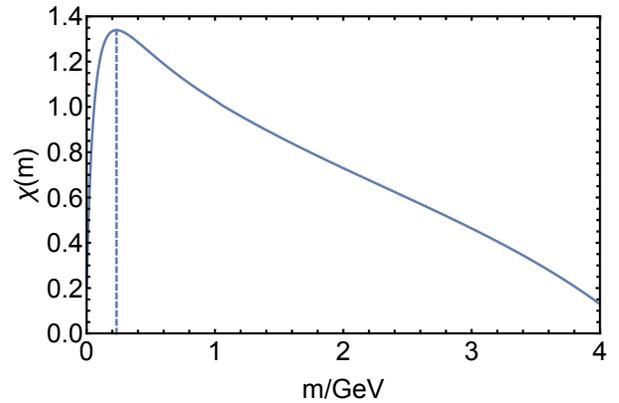}
\vspace*{-2mm}
\caption{The calculated dimensionless chiral susceptibility $\chi(m)$ where the dashed vertical line indicates the steepest descent mass of the corresponding condensate $-\langle\bar{q}q\rangle^{1/3}$.}
\label{fig:chic}
\end{figure}

The calculated current quark mass dependence of the dimensionless chiral susceptibility is shown in  Fig.~\ref{fig:chic}.
The response of the quark condensate to the current mass increases near chiral limit and then decreases gradually.
There then exists a critical mass at $m_{\rm crit}=0.22\;$GeV.
This critical mass reflects the relation between the DCSB and the ECSB effect through the interference.
For $m < m_{\rm crit}$, the $\chi(m)$ function is increasing, which indicates that the interference is mainly induced by the DCSB effect,
while for $m > m_{\rm crit}$, the $\chi(m)$ function is decreasing, indicating that the interference is mainly induced by the ECSB effect.
Therefore, the $\chi(m)$ function can be taken to identify which effect dominates the interference.

Also, the behavior of $\chi(m)$ is surprisingly consistent with the analysis of PDA of hadrons~\cite{PDA1,PDA2}.
For the mesons with light quark, the PDA is a broadening function in comparison to the asymptotic form $6x(1-x)$, while for heavy quarkonia, PDA becomes narrower.
The PDA of hadrons gives then this critical mass also in the neighborhood of the strange quark, where the respective PDA becomes precisely the asymptotic form.
This consistence reveals the connection between the internal hadron dynamics and  the condensate of quarks inside.
The different response of the quark condensate to the external source might induce an intrinsic transition of internal hadron dynamics.

Noticing that the condensate defined above is purely in terms of the quark propagator and thus independent of hadrons.
It is then interesting to investigate the behavior of the condensate when putting the quarks into hadrons.
In considering the pion, these features are expressed in the axial-vector Ward-Takahashi identity (AV-WTI)~\cite{MRT:1998,BC,CLR:2013}.
Equating pole contributions of the corresponding axial and pseudo-scalar vertexes in the AV-WTI, it entails
\begin{eqnarray}
f_{\pi}^2 m_{\pi}^2 = (m_u^{\zeta}+m_d^{\zeta}) f_{\pi} \rho_{\pi}^{\zeta},
\label{eq:kpi}
\end{eqnarray}
where $f_\pi$ and $\rho_\pi^\zeta$ are defined as:
\begin{equation} \label{eq:kpi}
\begin{split}
i f_{\pi} P_\mu=&\langle 0|\bar{q}\gamma_5\gamma_\mu q|\Pi \rangle \\
=&Z_2(\zeta,\Lambda)\trD\int_k^\Lambda i\gamma_5 \gamma_\mu S(k_+)\Gamma_\pi(k;P) S(k_-),
\end{split}
\end{equation}
\begin{equation} \label{eq:kpi}
\begin{split}
i \rho_\pi^\zeta=&-\langle 0|\bar{q}i\gamma_5 q|\Pi \rangle \\
=&Z_4(\zeta,\Lambda)\trD\int_k^\Lambda \gamma_5 S(k_+)\Gamma_\pi(k;P) S(k_-),
\end{split}
\end{equation}
respectively, where $\Gamma_\pi(k;P)$ is the pion Bethe-Salpeter amplitude, which reads
\begin{eqnarray}
\Gamma_\pi(k;P)=\int_q^\Lambda S(q_+)\Gamma_\pi(q;P) S(q_-) K(q,k;P),
\end{eqnarray}
where $K(q,k;P)$ is the fully-amputated quark-antiquark scattering kernel and $P^2=-m_\pi^2; k_\pm=k\pm P/2; q_\pm=q\pm P/2$, without loss of generality in a Poincar\'e covariant approach.

The relation connects the pion mass $m_{\pi}$ and decay constant $f_{\pi}$ to the $u$ and $d$ quark current masses and $\kappa_{\pi}^{\zeta} := f_{\pi} \rho_{\pi}^{\zeta}$,
namely the in-hadron condensate introduced in Refs.~\cite{MRoberts:1997,in-hadron:2012}.
The identity holds in the whole range of the current quark mass.

Using QCD's quark-level Goldberger-Treiman relations, one can prove~\cite{MRT:1998}
\begin{eqnarray}
f_{\pi}^0 \rho_{\pi}^{\zeta 0} = -\langle\bar{q}q\rangle^{\zeta 0},
\end{eqnarray}
where the superscript $0$ indicates that the quantity is computed
in the chiral limit. The relation means the two definitions of condensate coincide at chiral limit.
However, the in-hadron condensate could deviate against the quark condensate,
since the quarks are now in the bound state and the interaction could induce the change of  the  DCSB effect of quarks.
To identify such a point explicitly,  we define the ratio of the two condensates as:
\begin{eqnarray}  \label{eq:Rhm}
r_{h}(m):=\frac{f_{\pi} \rho_{\pi}}{ - \langle\bar{q}q\rangle},
\end{eqnarray}
\begin{figure}
\centering
\includegraphics[width=0.44\textwidth]{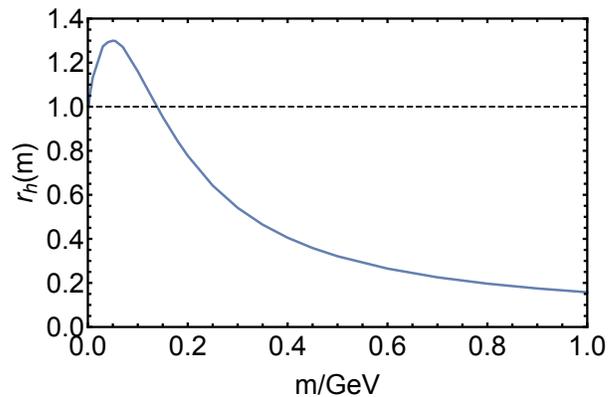}\;
\vspace*{-2mm}
\caption{The computed ratio $r_{h}(m)$ defined in Eq.~\eqref{eq:Rhm}, where the dashed horizontal line indicates $r_{h}(m)=1$. }
\label{fig:Rhm}
\end{figure}

The calculated current mass dependence of the ratio $r_{h}(m)$ is illustrated in Fig.~\ref{fig:Rhm}.
As we can see in Fig.~\ref{fig:Rhm}, in sufficiently small current masses with $r_{h} > 1 $,
the in-hadron condensate is larger than the quark condensate, which means that the interaction enhances the DCSB effect of quarks.
And again,  we can find a critical mass in the neighborhood of the strange quark, $m=0.13\;$GeV, after that,
the ratio is smaller than 1 where the in-hadron condensate starts to be smaller than the quark condensate.
In this region, the interaction in bound states reduces the DCSB effect and lead the system gradually to the perturbative region.
This again illustrates the transition of the internal hadron dynamics,
and moreover, it manifests that this transition is induced by the interference between the DCSB and the ECSB effects of the quark .

\section{Summary and outlook}
\label{SaO}

In summary, we have shown that a cut-off independent condensate can be extracted after analyzing the asymptotic behavior of the quark mass function.
This verifies a well-defined quark condensate at the level of quark propagator beyond chiral limit.
Along with the increase of current quark mass, the DCSB effect manifested via the condensate also shows an increasing behaviour
because of the interference of the mass scale between the two mass generation mechanisms.
The relative contribution can be evaluated after defining a dimensionless chiral susceptibility.
It gives then a critical mass in the neighborhood of the strange quark mass, $m=0.22\;$GeV.
When the current quark mass is smaller, the ECSB effect enhances the DCSB effect.
For heavier quarks, the ECSB effect reduces the DCSB effect which is then consistent with the NRQCD computations of heavy quarkonia.

Moreover, as we mentioned above, this condensate is independent of hadrons, and then different from the in-hadron condensate.
This difference reflects the effect of the interaction in hadron.
In fact, again we find a critical mass close to the value from dimensionless chiral susceptibility, $m=0.15\;$GeV.
For lighter quarks, as indicated by the dimensionless chiral susceptibility, the external source enhances the DCSB effect,
and consequently, the interaction in hadron brings in stronger DCSB effect.
When the quark mass exceeds the critical mass, the in hadron condensate is then smaller and thus the DCSB effect is reduced by the inside interaction of hadrons.

These observations are surprisingly consistent with the previous studies of PDA of hadrons.
As the current quark mass increases, the PDA  of mesons with equal quark mass changes from a broadening function to a narrow peak,
and there exists then a critical mass also  in the neighborhood of the strange quark, where the respective PDA becomes precisely the asymptotic form $6x(1-x)$.
All these phenomena reveal that an intrinsic transition of internal hadron dynamics takes place, and more interestingly,
there could  be signals from the experiments of strange matter since the location of the transition is very close.


\begin{acknowledgments}
The work was supported by the National Natural Science Foundation of China under Contracts No. 11435001 and No. 11775041.
%
%
And FG is grateful for the support from Alexander von Humbodlt Foundation.
\end{acknowledgments}

\end{document}